\documentclass[12pt,letterpaper]{article}
\usepackage{graphicx}
\usepackage{mathrsfs}
\def\gsim{\mathrel{\rlap{\lower4pt\hbox{\hskip1pt$\sim$}}\raise1pt\hbox{$>$}}}
\title{Running of Newton's Constant and Quantum Gravitational Effects\footnote{Invited contribution to the proceedings of the 46th course of the International School of Subnuclear Physics held in Erice, Sicily/Italy, from August 29 to September 7, 2008. Based on the talk given by the author in Erice and on extended versions presented elsewhere during fall 2008. For the original research work, see \cite{runningpaper,gutpaper}.}}\author{David Reeb\footnote{Electronic address:~\texttt{dreeb@uoregon.edu}}\\\normalsize{\textit{Institute of Theoretical Science, University of Oregon}}\\\normalsize{\textit{Eugene, OR 97403, USA}}}\date{}
\begin{document}
\maketitle
\thispagestyle{empty}
\begin{abstract}
Newton's gravitational constant $G_N$ is shown to be a running coupling constant, much like the familiar running gauge couplings of the Standard Model. This implies that, in models with appropriate particle content, the \emph{true} Planck scale, i.e.~the scale at which quantum gravity effects become important, can have a value different from $10^{19}\,{\rm GeV}$, which would be expected from naive dimensional analysis. Then, two scenarios involving this running effect are presented. The first one is a model which employs huge particle content to realize quantum gravity at the TeV scale in 4 dimensions, thereby solving the hierarchy problem of the Standard Model. Secondly, effects of the running of Newton's constant in grand unified theories are examined and shown to introduce new significant uncertainties in their predictions, but possibly also to provide better gauge coupling unification results in some cases.
\end{abstract}
\section{Running of Newton's constant}
\begin{figure}[tb]
\center{\includegraphics[scale=0.85]{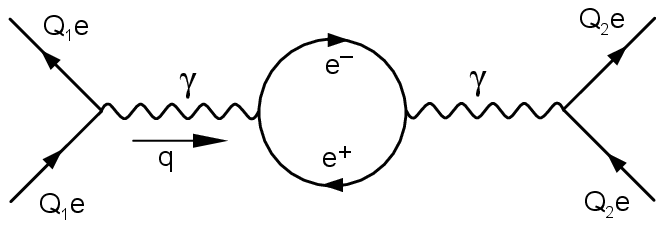}\hspace{13mm}\includegraphics[scale=0.63]{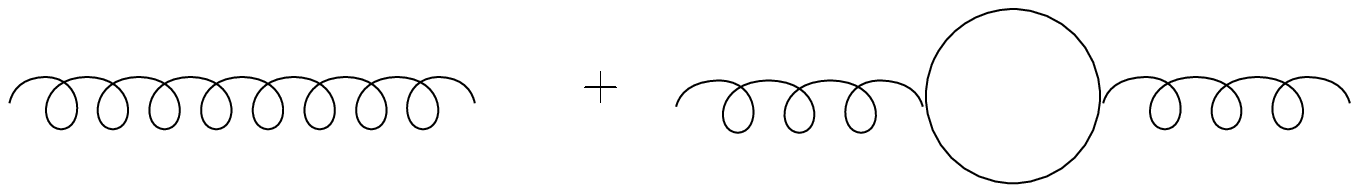}}
\caption{(left) The first-order contribution to virtual photon exchange in QED, having an electron-positron pair in the loop, modifies the effective gauge coupling of the photon to the external particles. (right) In a theory of quantized gravity, the effective graviton propagator is the sum of a tree-level contribution and loop diagrams, in particular with matter particles running in the loop.}
\label{loopdiagrams}
\end{figure}

We are familiar with the idea that the gauge couplings of the Standard Model of particle physics are running coupling constants, i.e.~that the values of these couplings depend on the energy scale at which we perform experiments or, equivalently, on the distance scale at which we probe a certain feature of Nature. The intuitive reason, in the case of the electromagnetic interaction (QED) for example, is that virtual ${\rm e}^+{\rm e}^-$ pairs polarize the vacuum and thus screen electric charges at large distances. Formally this comes about because, in an electromagnetic process with momentum transfer $q$ roughly greater than the electron mass $m_{\rm e}$, the tree-level diagram for virtual photon exchange receives significant superposition from the one-loop diagram with an electron in the loop, shown in Fig.~\ref{loopdiagrams}~(left). This modifies the photon propagator and hence effectively modifies the gauge couplings at the vertices, so that the effective coupling becomes dependent on energy scale $\alpha_{\rm EM}=\alpha_{\rm EM}(q)$:\begin{equation}
\label{gaugerunning}
\alpha_{\rm EM}(q)=\frac{\alpha_{\rm EM}(m_{\rm e})}{1-\frac{\alpha_{\rm EM}(m_{\rm e})}{3\pi}\log\left(q^2/m_{\rm e}^2\right)}~~~~~{\rm for}~q^2\gsim m_{\rm e}^2\,.
\end{equation}After Fourier transformation into real space this can be equivalently expressed by saying that, at distances $r$ shorter than the Compton wavelength of the electron $m_{\rm e}^{-1}$, the force between charges is bigger than expected by extrapolating Coulomb's law from large to small distances. The same screening or antiscreening behavior, depending on the particle content of the theory, occurs in non-Abelian gauge theories as well. So, a natural question to ask is whether similar effects might be present in gravity, especially in light of the fact that Newton's force law very much resembles Coulomb's law in form albeit with a dimensionful coupling constant, namely Newton's constant $G_N$.

\begin{figure}[htb]
\center{\includegraphics[width=8cm]{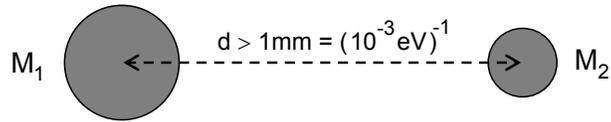}}
\caption{Our experiments measure Newton's constant $G_N$ only at macroscopic distances.}
\label{measureGN}
\end{figure}

When referring to Newton's constant, we usually implicitly think of the value $G_N=(10^{19}\,{\rm GeV})^{-2}$ in natural units $\hbar=c=1$, or similarly, in order to have a mass scale, we talk about the Planck mass $M_{\rm Pl}=G_N^{-1/2}=10^{19}\,{\rm GeV}$. But these values are only the outcomes of our macroscopic experiments, as shown in Fig.~\ref{measureGN}: basically, the force $F$ between two masses $M_1$ and $M_2$, separated by a distance $d$, is measured and $G_N$ is deduced via Newton's law $F=-G_NM_1M_2/d^2$. The smallest distances probed so far are on the order of $d\sim1\,{\rm mm}$ \cite{Hoyle:2004cw}, corresponding to an energy scale of $q\sim10^{-3}\,{\rm eV}$ --- very much smaller than energy scales at the current frontier in high-energy physics. For this reason it is sensible to indicate that $G_N(\mu\approx0\,{\rm GeV})=(10^{19}\,{\rm GeV})^{-2}$ and $M_{\rm Pl}(\mu\approx0\,{\rm GeV})=10^{19}\,{\rm GeV}$ are the values in the far infrared $\mu\approx0\,{\rm GeV}$. But how is this value of $G_N$ at infrared energies related to physics at high energy scales or, equivalently, to physics at short distances where gravity has not been probed yet?

I will first illustrate the scale dependence of Newton's constant in a crude way, using old-fashioned cutoff regularization. The action that describes gravity and matter
\begin{equation}
S_{\rm grav+m}=\int d^4x\sqrt{-\det g_{\mu\nu}}\left(\frac{1}{16\pi G_b}R(g_{\mu\nu})+\frac{1}{2}g^{\mu\nu}\partial_\mu\phi\partial_\nu\phi+\ldots+\psi+A_\mu+\ldots\right)
\end{equation}
constains the Einstein-Hilbert term, which describes the propagation and self-interaction of gravitons $g_{\mu\nu}$, and also couples all matter fields (scalars $\phi$, fermions $\psi$ and gauge bosons $A_\mu$) to gravity. As in the case of QED above, these couplings lead to corrections to the graviton propagator, Fig.~\ref{loopdiagrams}~(right), since matter particles can run in the loop. For momentum transfer $q$ and loop momentum cutoff $\Lambda$, the sum of both diagrams is, neglecting the index structure,
\begin{equation}
\label{Gren}
\frac{iG_b}{q^2}+\frac{iG_b}{q^2}\left(i\frac{c}{16\pi^2}q^2\Lambda^2\right)\frac{iG_b}{q^2}~=~\frac{iG_{\rm ren}}{q^2}~,
\end{equation}
where the loop contribution follows from dimensional analysis and is proportional to the square of the cutoff. If, as in QED, this one-loop contribution is absorbed into a redefined coupling constant $G_{\rm ren}$ as shown in (\ref{Gren}), the relation for the effective coupling is
\begin{equation}
\label{cutoffrunning}
\frac{1}{G_{\rm ren}}=\frac{1}{G_b}+\frac{c}{16\pi^2}\Lambda^2~.
\end{equation}
The appearance of the cutoff $\Lambda$ in (\ref{cutoffrunning}) qualitatively proves that Newton's constant is scale dependent and, moreover, that it depends \emph{quadratically} on energy, as opposed to only logarithmic running (\ref{gaugerunning}) in the case of the gauge coupling constants.

Regularization with a hard cutoff is very crude in the sense that it violates general coordinate invariance as well as gauge invariance (if gauge bosons in the loop are considered). Nevertheless, the above calculation can be done using heat-kernel regularization \cite{Larsen:1995ax,runningpaper} which is a method to integrate out (modes of) matter fields directly from the path integral in the presence of a gravitational background $g_{\mu\nu}$, yielding an effective action $S_{\rm eff}(g_{\mu\nu};\,\mu)$ for gravity appropriate to experiments performed at scale $\mu$. This method respects general covariance (and gauge invariance can be ensured with usual field-theoretic methods), so that the coefficient of $\sqrt{g}R$ in the effective gravitational action can be related to a running $G(\mu)$ in the ordinary Wilsonian sense. In the end, if $n_0$ real scalar fields, $n_{1/2}$ Weyl fermions and $n_1$ gauge bosons are integrated out, the values of Newton's constant at two different energy scales $\mu$ and $\mu_0$ are related via
\begin{equation}
\frac{1}{G(\mu)}=\frac{1}{G(\mu_0)}-\frac{\mu^2-\mu_0^2}{12\pi}\left(n_0+n_{1/2}-4n_1\right)~.
\label{wilsoneqn}
\end{equation}
This running very much resembles the running of gauge couplings (\ref{gaugerunning}) with the significant difference that, as was already obvious from (\ref{cutoffrunning}), here the coupling constant $G(\mu)$ depends quadratically on energy scale $\mu$, owing to its dimensionful nature.

If several matter fields are integrated out completely, the scale-dependent $G(\mu)$ can be expressed in terms of the measured low-energy $G(\mu_0=0)=G_N=(10^{19}\,{\rm GeV})^{-2}$ by
\begin{equation}
\label{allintegratedout}
\frac{1}{G(\mu)}=\frac{1}{G_N}-\mu^2\frac{N}{12\pi}~~~~~{\rm with}~\,N\equiv n_0+n_{1/2}-4n_1\,,
\end{equation}
where $N$ counts the field content integrated out. Clearly, having scalars or fermions in a theory increases the strength of gravity $G(\mu)$ at high energies $\mu>0$, whereas gauge bosons drive $G(\mu)$ in the opposite direction.

For the case of $N>0$ in particular, this scale-dependence of Newton's constant has one important consequence which will be the basic starting point for both applications in sections \ref{application1} and \ref{application2}: The Planck scale is defined to be the energy scale at which quantum gravitational effects become important. Naively, by dimensional analysis, this is the scale $M_{\rm Pl}=G_N^{-1/2}=10^{19}\,{\rm GeV}$; however, a running Newton's constant can change this picture. The \emph{true} Planck scale $\mu_*$ is given by the condition
\begin{equation}
\label{defplanckscale}
\mu_*=G(\mu_*)^{-1/2}~,
\end{equation}
i.e.~$\mu_*$ is the energy scale so that fluctuations in spacetime geometry at the corresponding length scales $<\mu_*^{-1}$ are unsuppressed. $\mu_*$ is the true Planck scale, whereas our value $M_{\rm Pl}=10^{19}\,{\rm GeV}$ should be thought of as derived from this fundamental $\mu_*$ via renormalization group running to very low energies. Condition (\ref{defplanckscale}) yields, with (\ref{allintegratedout}),
\begin{equation}
\label{mustar}
\mu_*=\frac{M_{\rm Pl}}{\sqrt{1+N/12\pi}}=\frac{M_{\rm Pl}}{\eta}~~~~~{\rm with}~\,\eta\equiv\sqrt{1+N/12\pi}\,,
\end{equation}
where $N/12\pi$ in the denominator is the effect of the one-loop diagram in Fig.~\ref{loopdiagrams} (right) with particle content $N$ in the loop, correcting the tree result $1$. $N$, or equivalently $\eta$, characterizes the matter content of a certain theory. Gravitons (through self-coupling) will also run in the loop and give an additional contribution, but its value is fixed and comparable in size to the running caused by one matter field; and in the applications in the following two sections we are interested in the behavior of $\mu_*$ as we vary the matter content $N$ so that an additional small constant contribution would only affect unimportant details.
\section{TeV quantum gravity in 4 dimensions}\label{application1}
\begin{figure}[ht]
\center{\includegraphics[width=90mm,height=70mm]{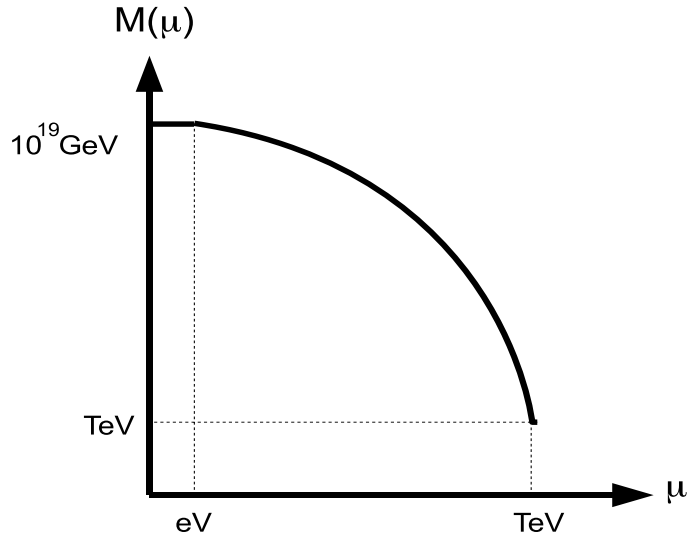}}
\caption{Running of the gravitational mass scale $M(\mu)=G(\mu)^{-1/2}$ in a theory with particle content $N\sim10^{32}$ from the fundamental scale $\mu_*={\rm TeV}$ to our low-energy value $M_{\rm Pl}=10^{19}\,{\rm GeV}$ at $\mu<{\rm eV}$.}
\label{mrunning}
\end{figure}

In this section, I will present a model, in four spacetime dimensions, in which the running of Newton's constant pushes the Planck scale all the way down to the weak scale, i.e.~to the TeV region, despite the conventional wisdom according to which quantum gravity effects are weak at energies accessible to us.
With formula (\ref{mustar}) for the true scale of quantum gravity at hand, it is natural to ask what the particle content of a theory has to be in order to have a fundamental gravity scale of $\mu_*\sim {\rm TeV}$; at this scale then new quantum gravitational physics would have to appear, and it would coincide with the scale at which the Standard Model of particle physics becomes finely tuned und likewise with the scale that will be accessible to our particle accelerators in the very near future. From (\ref{mustar}), this condition $\mu_*\sim{\rm TeV}$ implies a particle content of $N\sim 10^{32}$. I.e., if a theory contains $\sim 10^{32}$ scalars or fermions with masses below a TeV (so that they can run in the loop), then our measured low-energy value of Newton's constant would be consistent with a fundamental quantum gravity scale of $\mu_*\sim{\rm TeV}$. This behavior, translated into a running gravitational mass scale $M(\mu)\equiv G(\mu)^{-1/2}$, is illustrated in Fig.~\ref{mrunning}, where the fundamental value $M({\rm TeV})={\rm TeV}$ is fixed and all lower values are obtained via running (\ref{wilsoneqn}), in particular also the value $M({\rm eV})=10^{19}\,{\rm GeV}$ that we observe in the macroscopic world.

Thus, a model which contains $10^{32}$ scalars or fermions with masses $m<{\rm TeV}$ interacting only gravitationally with us could have quantum gravity at a TeV, and yet be consistent with our low-energy measurements. To be phenomenologically viable, this vast amount of new fields has to be contained in a sector hidden from the Standard Model, i.e.~must not have any gauge or other interactions with us. Note that such a model could be supersymmetric without contradiction since, according to (\ref{allintegratedout}), the bosonic and fermionic components of a chiral supermultiplet give contributions of the same sign to the running of $G(\mu)$. This mechanism solves the hierarchy problem of the Standard Model since new physics describing fluctuations in spacetime geometry (i.e., quantum gravity) then would have to come in at the TeV scale and would cut off the Standard Model there. In this scenario, quantum gravity could be probed by experiments at the CERN LHC. Also, this model is the first one to feature quantum gravity in the TeV region already in four dimensions, which was previously only known to be possible with extra-dimensional models. And although the introduction of $\sim 10^{32}$ hidden particles in our model seems to be very drastic and unattractive compared to the proposition of a few extra dimensions, there is a connection between the two scenarios closely linked to this huge number $10^{32}$, as will be shown now.

In a model with $(d-4)$ extra dimensions, the fundamental action
\begin{equation}
\label{xdimaction}
S=\int d^4x\,d^{d-4}x'\sqrt{-g_{(d)}}\left(M_{(d)}^{d-2}{\cal R}+\ldots\right)~\equiv~\int d^4x\sqrt{-g_{(4)}}\left(M_{(4)}^2R+\ldots\right)
\end{equation}
is an integral over $d$-dimensional space with metric $g_{(d)}=g_{(4)}g_{(d-4)}$ involving the fundamental gravitational mass scale $M_{(d)}$ in $d$ dimensions. Since we, confined to a 4-dimensional brane, want to describe physics by a 4-dimensional action integral, we choose the description on the RHS of (\ref{xdimaction}). The mass scale $M_{(4)}$ that seems to be fundamental to 4-dimensional beings is, as per (\ref{xdimaction}), merely induced by more fundamental quantities in the $d$-dimensional world:
\begin{equation}
\label{volumetrick}
M_{(4)}^2=\int d^{d-4}x'\sqrt{g_{(d-4)}}\,M_{(d)}^{d-2}=V_{d-4}M_{(d)}^{d-2}~,
\end{equation}
with the proper volume $V_{d-4}$ of the extra $(d-4)$ dimensions. Now we, in the 4-dimensional world, are wondering why our observed value $M_{(4)}=M_{\rm Pl}=10^{19}\,{\rm GeV}$ is so much larger than the mass scales in all other theories we know (e.g., weak scale). Extra-dimensional models explain this discrepancy by noting that, according to (\ref{volumetrick}), $M_{(4)}$ can be large despite having a comparatively small fundamental scale of $M_{(d)}\sim {\rm TeV}$ if only the (unaccessible) volume $V_{d-4}$ of the extra dimensions is large. But the number of degrees of freedom hidden in the $(d-4)$-dimensional bulk equals the number of fundamental Planck volumes $V_{\rm Pl}$ in $V_{d-4}$, i.e.~if one wants to achieve $M_{(d)}\sim{\rm TeV}$ this number invariably is, with (\ref{volumetrick}),
\begin{equation}
V_{d-4}/V_{\rm Pl}=V_{d-4}M_{(d)}^{d-4}=M_{(4)}^2/M_{(d)}^2\sim\left(10^{19}\,{\rm GeV}/{\rm TeV}\right)^2\sim10^{32}~.
\end{equation}
These $10^{32}$ hidden degrees of freedom in the bulk are just as unaccessible as the $10^{32}$ bosonic or fermionic degrees of freedom in our model above, and this huge number is simply dictated, as in (\ref{mustar}), by the square of the ratio of scales. Viewed under this perspective, our model does not seem to be much more ugly than extra dimensions.

The most striking phenomenological feature of our model is that it exhibits strong gravity effects at the scale of a TeV. Quantum black holes might be created at the LHC with a (geometrical) cross section as large as $\sigma(pp\to qBH+X)\sim10^5\,{\rm fb}$ \cite{runningpaper} and would thus dominate all Standard Model cross sections. There is a sizable cross section $\sigma(pp\to {\rm m.e.})\sim750\,{\rm fb}$ into missing energy (whereby black holes decay into the $10^{32}$ hidden fields) which seems to be larger than in extra-dimensional RS- or ADD-models in which almost all energy remains on the brane \cite{Calmet:2008dg}. The quantum black holes created are likely charged under ${\rm SU(3)}_{\rm c}\times{\rm U(1)}_{\rm Q}$ since they are created from colored and possibly electromagnetically charged partons (color is not confining at such small distances). Thus, although the number of hidden fields by far exceeds the number of Standard Model species, decay of these quantum black holes involving one or two Standard Model particles is likely due to charge conservation, and one particular characteristic signature is a back-to-back hard lepton+jet \cite{Calmet:2008dg}.

In conclusion, the enormous number of particles required below a TeV, which must interact only gravitationally with us, seems unattractive, and other issues, like early cosmology above a TeV, pose difficulties and have not been addressed. Regardless which model might achieve strong gravity at the TeV scale, if any, it at least seems to require drastic means to realize this possibility, be it either the assumption of extra dimensions or the introduction of a large number of extra particles.

A slightly different possibility to overcome the large hierarchy between the weak scale and $M_{\rm Pl}=10^{19}\,{\rm GeV}$ might be modifications to gravity itself: the coefficients of the higher-derivative terms in the gravitational action
\begin{equation}
S=\int d^4x\sqrt{-g}\left(c_1R+c_2R^2+c_3R_{\mu\nu}R^{\mu\nu}+\ldots\right)
\end{equation}
are very weakly constrained, $c_2,\,c_3<10^{61}$ \cite{runningpaper}. It is thus conceivable that boundary values $c_i(\mu)$ at the scale $\mu_*\sim{\rm TeV}$ exist which through large gravitational self-couplings lead to the observed huge value of $c_1(\mu=0)=M_{\rm Pl}^2/16\pi$. Incidentally, coefficients $c_2,\,c_3\sim\left(10^{32}\right)^2$ would be expected after integrating out $10^{32}$ matter fields.
\section{Quantum gravitational effects on grand unification}\label{application2}
As a second scenario, I want to elucidate what effects the running of Newton's constant can have in more mundane models without unusually large particle content. In particular, this section examines how quantum gravity and running $G_N$ affect grand unified theories, especially supersymmetric ones. Results could influence whether one considers evidence for, or the possibility of, grand unification as plausible, and might also impact expectations of what we will find at the LHC.

\begin{figure}[tb]
\center{\includegraphics[height=6.5cm,width=14cm]{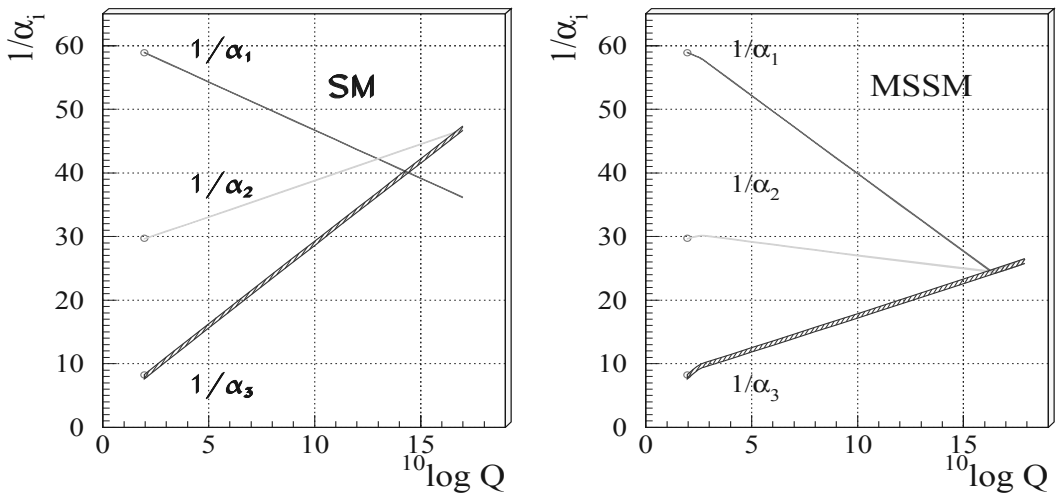}}
\caption{The inverses of the three gauge coupling constants evolved from their precisely measured values at $M_Z$ to higher energies using the renormalization group equations of the Standard Model (left panel) and of the Minimal Supersymmetric Standard Model with SUSY breaking scale $M_{\rm SUSY}\sim1\,{\rm TeV}$ (right).}
\label{unificationfig}
\end{figure}

The gauge coupling constants of the Standard Model are measured very accurately at the (low) energy scale $M_Z$ (mass of the $Z$ boson), for example by the LEP experiment: $\alpha_1(M_Z)=0.016887$, $\alpha_2(M_Z)=0.03322$, $\alpha_3(M_Z)=0.118(5)$ \cite{pdg}. If these couplings are evolved to higher energies using the renormalization group evolution equations of the Standard Model, the couplings run towards a common region, but do not quite meet (Fig.~\ref{unificationfig}, left panel). However, if they are evolved with the RG equations of the Minimal Supersymmetric Standard Model using a phenomenologically plausible SUSY breaking scale of $M_{\rm SUSY}\sim1\,{\rm TeV}$, then the measured low-energy couplings seem to be just right for the evolved couplings to meet at some common value near a unification scale of $M_X=10^{16}\,{\rm GeV}$. This meeting is pretty much exact if two-loop beta functions are used (right panel of Fig.~\ref{unificationfig}), which now has become part of the standard analysis \cite{Amaldi:1991cn}. The fact that the numerical values of the three gauge couplings meet under SUSY RG evolution at a high energy scale is often taken is strong evidence for supersymmetric grand unification. I.e.~on the one hand it is taken as evidence for grand unification, the idea that the three gauge couplings arise from one coupling of a unified simple gauge group which is spontaneously broken at $M_X$ via Higgs mechanism; and on the other hand it is also often quoted as evidence for SUSY itself, which would then be expected to be found soon at the LHC.

How do quantum gravity effects influence this picture? First, the alleged unification scale of $M_X=10^{16}\,{\rm GeV}$ is uncomfortably close to the naive Planck scale $M_{\rm Pl}=10^{19}\,{\rm GeV}$, and the \emph{true} Planck scale $\mu_*$ might be even lower (\ref{mustar}). Implicit in the above description of grand unification was the unification condition $\alpha_1(M_X)=\alpha_2(M_X)=\alpha_3(M_X)$ of numerical equality in the values of the evolved couplings, and it is this condition that is changed by quantum gravity.

Quantum gravity induces (non-perturbatively) a dimension-5 operator in the (effective) grand unified theory \cite{Hill:1983xh,Shafi:1983gz}
\begin{equation}
\label{dim5operator}
{\mathscr L}~\sim~\frac{c}{\hat{\mu}_*}{\rm Tr}\left(G_{\mu\nu}G^{\mu\nu}H\right)~~~~~{\rm with}~\,c\sim{\cal O}(1)\,,
\end{equation}
which is suppressed by the reduced true Planck scale $\hat{\mu}_*=\mu_*/\sqrt{8\pi}$, the quantity that enters quantum gravity calculations, and so naturally has dimensionless coefficient $c$ of order 1. $G_{\mu\nu}$ is the gauge field strength of the grand unified theory and $H$ is one of the Higgs fields needed to break the unified gauge group down to the Standard Model group ${\rm SU(3)}\times{\rm SU(2)}\times{\rm U(1)}$. Below the scale $M_X$ of grand unified symmetry breaking the Higgs field acquires a vacuum expectation value $\langle H\rangle$, and with the replacement $H\to\langle H\rangle$ the operator (\ref{dim5operator}) modifies the gauge kinetic term of the Standard Model vector bosons which are massless below $M_X$:
\begin{equation}
{\mathscr L}~\sim~-\frac{1}{2}{\rm Tr}\left(G_{\mu\nu}G^{\mu\nu}\right)+\frac{c}{\hat{\mu}_*} {\rm Tr}\left(G_{\mu\nu}G^{\mu\nu}\langle H\rangle\right)~\sim~\sum_{i=1}^3-\frac{1}{2}\left(1+\epsilon_i\right){\rm Tr}\left(F^{(i)}_{\mu\nu}F^{(i)\mu\nu}\right)~,
\end{equation}
where $i$ runs over the factors of the Standard Model gauge group and the $\epsilon_i$ are calculable in each specific model of GUT breaking. In the case of minimal SU(5) unification, which we assume for now for simplicity, the multiplet $H$ in the adjoint representation acquires a vacuum expectation value $\langle H\rangle=M_X(2,2,2,-3-3)/\sqrt{50\pi\alpha_G}$, where $\alpha_G$ is the value of the SU(5) gauge coupling at $M_X$, so that
\begin{equation}
\epsilon_1=\frac{\epsilon_2}{3}=-\frac{\epsilon_3}{2}=\frac{\sqrt{2}}{5\sqrt{\pi}}\frac{c\eta}{\sqrt{\alpha_G}}\frac{M_X}{\hat{M}_{\rm Pl}}~.
\end{equation}
Thus, the three gauge kinetic terms of the Standard Model are modified in \emph{different} ways. After a finite field redefinition $A_{\mu}^i\to(1+\epsilon_i)^{1/2}A_{\mu}^i$ the kinetic terms are canonically normalized below $M_X$, and it is then the corresponding redefined coupling constants $g_i\to(1+\epsilon_i)^{-1/2}g_i$ that are observed at low energies and that obey the usual RG equations below $M_X$, whereas it is the \emph{original} coupling constants that need to meet at $M_X$ in order for unification to happen. In terms of the observable rescaled couplings, the unification condition therefore reads
\begin{equation}
\label{unificationcondition}
(1+\epsilon_1)\alpha_1(M_X)=(1+\epsilon_2)\alpha_2(M_X)=(1+\epsilon_3)\alpha_3(M_X)=\alpha_G~.
\end{equation}
It is worth emphasizing that there are two separate effects from quantum gravity playing in here: first, the presence of the gravitationally induced operator (\ref{dim5operator}) influences grand unification simply because $M_X$ is close to $M_{\rm Pl}$; and second, the \emph{true} scale of quantum gravity might be even lower due to running $G_N$ so that the dimension-5 operator is effectively enhanced by the factor $\eta$ from (\ref{mustar}), which is what our work focuses on \cite{gutpaper}.

Supersymmetric grand unified theories, which can seemingly better accommodate coupling constant unification (Fig.~\ref{unificationfig}), have typically quite large particle content $N$ (or equivalently $\eta$) because most of their fields are Higgses in chiral supermultiplets, which contain scalars and fermions giving contributions of the same sign to $\eta$. Also, grand unified groups larger than SU(5) are commonly considered since they can satisfy phenomenological constraints such as proton decay, R parity and fits to the mass spectrum more easily. In the minimal SUSY-SU(5) model with ${\bf 24}$, ${\bf 5}$ and ${\bf \overline{5}}$ Higgs multiplets the enhancement factor is $\eta=2.3$, but the ``minimal SUSY-SO(10) model'' with ${\bf 126}$, ${\bf \overline{126}}$, ${\bf 210}$ and ${\bf 10}$ Higgses already has $\eta=6.2$. Other models like ${\rm E8}\times{\rm E8}$ with ${\bf 248}$ and ${\bf 3875}$ Higgs multiplets clearly have even bigger $\eta$, such that $\hat{\mu}_*$ will be significantly lower than $\hat{M}_{\rm Pl}$. Since minimal SUSY-SU(5) is phenomenologically ruled out, $\eta\gsim5$ is a reasonable assumption for realistic SUSY GUTs.

I will now first illustrate the size of the corrections coming from the modified unification condition (\ref{unificationcondition}) by comparing it to the size of two-loop corrections in the running of gauge couplings, which have become part of the standard analysis of grand unification. In a toy model of SUSY-SO(10) one has $\eta=6.2$ and, with (\ref{unificationcondition}), SU(5)-like gauge symmetry breaking then requires a splitting between the numerical values $\alpha_i(M_X)$ of, for example,
\begin{equation}
\frac{\alpha_3(M_X)-\alpha_2(M_X)}{\alpha_3(M_X)}~\approx~\epsilon_2(c\eta) - \epsilon_3(c\eta) ~\approx~\left\{\begin{array}{l}+9\%~~~{\rm for}~c=+1\\-9\%~~~{\rm for}~c=-1.\end{array}\right.
\end{equation}
This is a big uncertainty in our knowledge of the correct unification condition since within the grand unified theory nothing can be inferred about the exact size $c$ of the induced operator, except that it is reasonable to assume $c\sim{\cal O}(1)$. In particular, even making this reasonable assumption, the uncertainty of $\pm9\%$ is much bigger than the corrections to the $\alpha_i(M_X)$ coming from the inclusion of two-loop running. These are smaller than 3.5\% for all $i=1,\,2,\,3$. Thus, two-loop computations do not improve evidence for grand unification, as is often claimed, since the size of, and the uncertainty in, the effects from quantum gravity is far greater than two-loop contributions.
\begin{figure}[tb]
\center{\includegraphics[width=115.5mm,height=80mm]{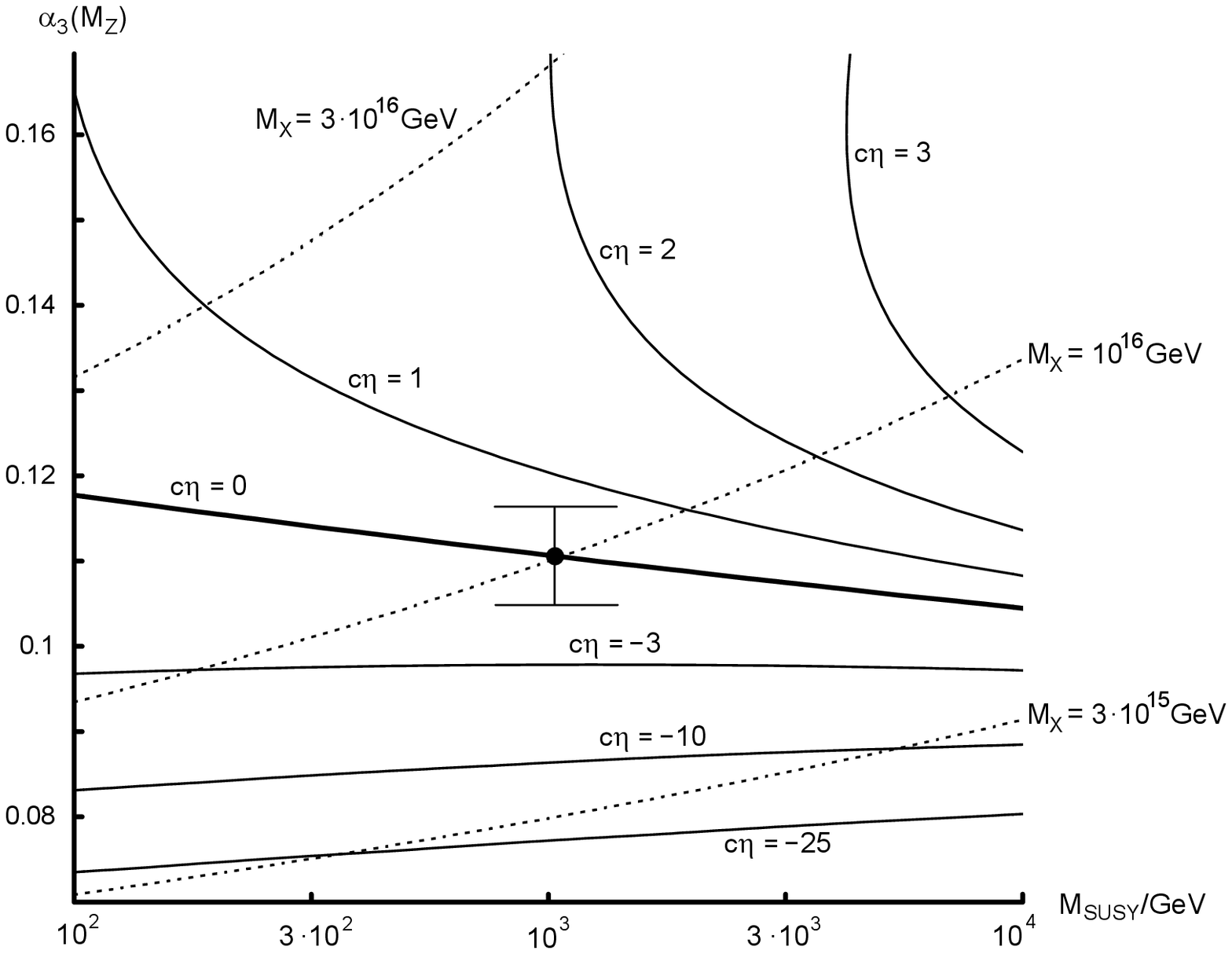}}
\caption{For $\eta$ fixed by the particle content of the theory, solid lines are contours of constant $c$ such that, under the modified unification condition (\ref{unificationcondition}), SUSY-SU(5) perfectly unifies at two loops for given values of the initial strong coupling $\alpha_3(M_Z)$ and SUSY breaking scale $M_{\rm SUSY}$.}
\label{contourplot}
\end{figure}

But not only is two-loop running of the gauge couplings too accurate in view of such large uncertainties from quantum gravity, also do claims of grand unification based on the pretty precise measurements of the low-energy couplings $\alpha_i(M_Z)$ not hold up in this light. For illustration, Fig.~\ref{contourplot} shows different low-energy inputs of the SUSY breaking scale $M_{\rm SUSY}$ and the strong coupling $\alpha_3(M_Z)$ at the $Z$ mass, while the relatively precisely measured $\alpha_1(M_Z)=0.016887$ and $\alpha_2(M_Z)=0.03322$ are held fixed; $M_{\rm SUSY}$ is expected to be in the region $10^{3\pm1}\,{\rm GeV}$ for phenomenological reasons, and the current uncertainty in $\alpha_3(M_Z)$ is indicated by a vertical bar. For each point in the diagram, these low-energy inputs have been evolved to higher energies with the supersymmetric two-loop renormalization group equations, and for each input it has been found that the modified unification condition (\ref{unificationcondition}) is satisfied for exactly one pair of values of the coefficient $c$ and $M_X$. Contour lines for these variables are shown in the figure, and $\eta\gsim5$ is imagined to be fixed by the particle content of the model. The coefficient $c$, which is only determined by quantum gravity, naturally ranges between $-1<c<1$, which covers a pretty large area in Fig.~\ref{contourplot}. In particular, for many of those values $c$, grand unification is incompatible with our measurements of low-energy parameters like $\alpha_3(M_Z)$: slight variations in $c$ would require unacceptably large adjustments in low-energy inputs $\alpha_i(M_Z)$ and $M_{\rm SUSY}$ for unification to still happen.

On the other hand, if $\eta\sim5$ then for a large range of low-energy inputs one can easily find a natural $c\sim{\cal O}(1)$ such that unification happens; under this perspective, unification does not seem very special. Our ignorance of $c\eta$ is far greater than the uncertainty in LEP measurements. And disturbingly, as can be seen from the figure, if the underlying theory of quantum gravity determines the coefficient of the operator (\ref{dim5operator}) to be $c>4/\eta\sim1$, then unification is impossible even far beyond the allowed SM/MSSM parameter range. The commonly quoted evidence for unification (right panel of Fig.~\ref{unificationfig}) arises because the central value of our measurements of low-energy parameters (black dot in Fig.~\ref{contourplot}) happens to lie almost perfectly on the contour line $c=0$. But considering the uncertainties, it seems that we cannot either suggest or rule out grand unification with any high degree of confidence based on low-energy observations alone, and thus the idea of grand unification loses much of its beauty. In particular, previous attempts to pin down $\alpha_3(M_Z)$ or $\sin^2\theta_W$ based on the assumption of unification seem invalid. Also, expectations about whether we will find supersymmetry at the LHC might need to be reconsidered if one previously regarded Fig.~\ref{unificationfig} as strong evidence for SUSY.

But one can also turn this reasoning around and argue that the gravitational operator (\ref{dim5operator}) with $c\sim{\cal O}(1)$ might facilitate gauge coupling unification in models where this did not seem to happen previously but which otherwise are phenomenologically viable. The picture in this case would be that the underlying fundamental theory leads to an effective grand unified theory with (strong) remnants from quantum gravity.
\section{Conclusion}
I have shown that Newton's constant is a running coupling constant, as are the familiar gauge couplings of the Standard Model, and that therefore the strength of gravity depends on energy scale. The value $G_N=(10^{19}\,{\rm GeV})^{-2}$ that we measure in the infrared is not directly related to the fundamental scale of quantum gravity, but is derived via renormalization group running from the fundamental value $\mu_*$. The exact relationship depends on the particle content of the theory under consideration, eqn.~(\ref{mustar}).

A four-dimensional model was presented that contains $\sim10^{32}$ scalars or fermions with masses below a TeV in a hidden sector, interacting with the Standard Model only gravitationally. This model solves, already in four dimensions, the hierarchy problem of the Standard Model since the large particle content implies that the fundamental scale of quantum gravity is around a TeV; at this scale then, new (gravitational) physics is expected to come in and replace the Standard Model. Popular extra-dimensional models were shown to postulate an equally large amount of degrees of freedom hidden in their bulk, and some phenomenology of our model was outlined.

In a second scenario, it was noted that in popular supersymmetric grand unified theories the true Planck scale is significantly lowered, mainly because of their large Higgs content. This amplifies the uncertainty in the size of operators that are induced by quantum gravity and so enhances uncertainties in unification predictions due to a modified unification condition. We have shown that this effect is much bigger than measurement uncertainties and the two-loop corrections commonly considered. In particular, our analysis might impact whether one considers \emph{apparent} unification of gauge couplings to be strong evidence for grand unification or SUSY. On the other hand, phenomenologically viable GUT models that are only lacking sufficient gauge coupling unification might be revived by the presence of such induced operators along with the running of Newton's constant.

\bigskip
\emph{Acknowledgments ---} I would like to thank Antonino Zichichi and Gerardus 't Hooft for organizing a wonderful 46th Course of the International School of Subnuclear Physics 2008, bringing together students and well-known researchers in a hospitable atmosphere in Erice, and for the opportunity to present this work there. Also with great pleasure I acknowledge collaboration with Xavier Calmet and Stephen Hsu. The original research work and the author's travel to Erice were supported by the Department of Energy under DE-FG02-96ER40969.


\begin{thebibliography}{99}
\bibitem{runningpaper}
X.~Calmet, S.~D.~H.~Hsu and D.~Reeb, Phys.\ Rev.\  D {\bf 77}, 125015 (2008) [arXiv:0803.1836 [hep-th]].
\bibitem{gutpaper}
X.~Calmet, S.~D.~H.~Hsu and D.~Reeb, Phys.\ Rev.\ Lett.\  {\bf 101}, 171802 (2008) [arXiv:0805.0145 [hep-ph]].
\bibitem{Hoyle:2004cw}
C.~D.~Hoyle \emph{et al.}, Phys.\ Rev.\  D {\bf 70}, 042004 (2004) [arXiv:hep-ph/0405262].
\bibitem{Larsen:1995ax}
F.~Larsen and F.~Wilczek, Nucl.\ Phys.\  B {\bf 458}, 249 (1996) [arXiv:hep-th/9506066].
\bibitem{Calmet:2008dg}
X.~Calmet, W.~Gong and S.~D.~H.~Hsu, Phys.\ Lett.\  B {\bf 668}, 20 (2008) [arXiv:0806.4605 [hep-ph]].
\bibitem{pdg}
W.--M.~Yao \emph{et al.}~(Particle Data Group), J.\ Phys.\ G {\bf 33}, 1 (2006).
\bibitem{Amaldi:1991cn}
U.~Amaldi, W.~de Boer and H.~Furstenau, Phys.\ Lett.\  B {\bf 260}, 447 (1991).
\bibitem{Hill:1983xh}
C.~T.~Hill, Phys.\ Lett.\  B {\bf 135}, 47 (1984).
\bibitem{Shafi:1983gz}
Q.~Shafi and C.~Wetterich, Phys.\ Rev.\ Lett.\  {\bf 52}, 875 (1984).
\end{thebibliography}
\end{document}